\documentclass[aps,preprint,showpacs,amsmath,superscriptaddress]{revtex4}

\usepackage{times}
\usepackage{hyperref}

\usepackage{graphicx}% Include figure files
\usepackage{dcolumn}% Align table columns on decimal point
\usepackage{bm}% bold math

\begin{document}

\title{Precision radiative corrections to the semileptonic Dalitz plot with angular correlation between polarized decaying and emitted
baryons: Effects of the four-body region}

\author{J.\ J.\ Torres
}
\affiliation{
Escuela Superior de C\'omputo del IPN, Apartado Postal 75-702, M\'exico, D.F.\ 07738, M\'exico
}

\author{M.\ Neri
}
\affiliation{
Escuela Superior de F{\'\i}sica y Matem\'aticas del IPN, Apartado Postal 75-702, M\'exico, D.F.\ 07738, Mexico
}

\author{A.\ Mart{\'\i}nez
}
\affiliation{
Escuela Superior de F{\'\i}sica y Matem\'aticas del IPN, Apartado Postal 75-702, M\'exico, D.F.\ 07738, Mexico
}

\author{A.\ Garc{\'\i}a
}
\affiliation{
Departamento de F{\'\i}sica, Centro de Investigaci\'on y de Estudios Avanzados del IPN, Apartado Postal 14-740, M\'exico, D.F.\ 07000,
Mexico
}

\author{Rub\'en Flores-Mendieta
}
\affiliation{Instituto de F{\'\i}sica, Universidad Aut\'onoma de San Luis Potos{\'\i}, \'Alvaro Obreg\'on 64, Zona Centro, San Luis
Potos{\'\i}, S.L.P.\ 78000, Mexico
}

\date{\today}

\begin{abstract}
Analytical radiative corrections of order $(\alpha/\pi)(q/M_1)$ are calculated for the four-body region of the Dalitz plot of baryon
semileptonic decays when the ${\hat {\mathbf s}_1} \cdot {\hat {\mathbf p}_2}$ correlation is present. Once the final result is
available, it is possible to exhibit it in terms of the corresponding final result of the three-body region following a set of simple
changes in the latter. We cover two cases, a charged and a neutral polarized decaying baryon.
\end{abstract}


\maketitle

In this report we shall improve the precision of analytical radiative corrections (RC) to the four-body region (FBR) of the Dalitz plot
of baryon semileptonic decays, $A\to Bl\overline{\nu}_l$, with a polarized decaying baryon by including the order $(\alpha/\pi)(q/M_1)$
contributions. The ${\hat {\mathbf s}_1} \cdot {\hat {\mathbf p}_2}$ angular correlation will be kept explicitly. The corresponding
corrections to order $(\alpha/\pi)(q/M_1)^0$ were obtained in Ref.~\cite{uno}. Here we follow the same notation and conventions, and
we recall that in this region neither the neutrino nor the real photon can reach zero energy. As explained there, this region must be
incorporated into the RC when bremsstrahlung photons cannot be discriminated either kinematically or by direct detection. This is the
case when only the emitted baryon momentum is measured and its corresponding asymmetry coefficient is determined.

The result we want cannot be obtained from the corresponding final result for the three-body region (TBR), because the variable $y_0$
plays a double role in the latter case. It appears in the integrand and as the upper limit of the $y={\hat {\mathbf p}_2} \cdot {\hat {\mathbf l}}$
integration variable [see Eqs.~(6) and (7) of Ref.~\cite{uno}]. In the FBR case $y_0$ is no more an upper limit of $y$, since now the
latter has one as an upper limit, and $y_0$ appears only in the integrand. One is then forced to perform the complete calculation.
However, once the final result is obtained, it is possible to present it in terms of the final result of the TBR by making simple
changes in the latter. This allows us to make a concise presentation and avoid many unnecessary repetitions.

The calculation of RC in the FBR has the same structure as in the TBR, so there is an exact parallelism all along. To mark the
difference we shall introduce a subindex $F$ in the expressions and definitions that pertain to the FBR. In this report we shall
present only new results, all others will be appropriately referenced.

The complete Dalitz plot without the restriction of kinematically discriminating real photons is
\begin{equation}
d\Gamma_i = d\Gamma_i^{\textrm{TBR}}+d\Gamma_i^{\textrm{FBR}}, \qquad i= \textrm{C,N}
\end{equation}
where $i=\textrm{C,N}$ refers to charged and neutral decaying baryon cases, respectively. The analytical RC to order $(\alpha/\pi)(q/M_1)$
to the TBR part including the ${\hat {\mathbf s}_1} \cdot {\hat {\mathbf p}_2}$ correlation are found in Eqs.~(24) and (44) of
Ref.~\cite{dos} for $i=\textrm{C,N}$, respectively. The FBR arises only from bremsstrahlung and it can be separated into an unpolarized
part $({\hat {\mathbf s}_1}=\mathbf{0})$ and into another one containing the ${\hat {\mathbf s}_1} \cdot {\hat {\mathbf p}_2}$
correlation, that is,
\begin{equation}
d\Gamma_i^{\textrm{FBR}}=d\Gamma_{iB}^{\prime\, \textrm{FBR}}-d\Gamma_{iB}^{(s)\,\textrm{FBR}}.
\end{equation}
Here the subindex $B$ stresses the bremsstrahlung origin of these parts. To order $(\alpha/\pi)(q/M_1)$ the spin-independent part is
found, with some changes in notation, in Eq.~(32) of Ref.~\cite{tres} for $i=\textrm{C}$ and in Eq.~(22) of Ref.~\cite{cuatro} for
$i=\textrm{N}$. To order $(\alpha/\pi)(q/M_1)^0$ the spin-dependent part is found in Eq.~(18) of Ref.~\cite{uno} for $i=\textrm{C,N}$.

To obtain the order $(\alpha/\pi)(q/M_1) $ of $d\Gamma_{\textrm{C} B}^{(s)\, \textrm{FBR}}$ we trace a parallelism with Eq.~(14) of
Ref.~\cite{dos} for the TBR. The final result is compactly given by
\begin{equation}
d\Gamma_{\textrm{C}B}^{(s)\, \textrm{FBR}}= \frac{\alpha}{\pi} d\Omega {\hat {\mathbf s}_1} \cdot {\hat {\mathbf p}_2} \left[
B_2^\prime I_{C0F} + C_{AF}^{(s)} \right]. \label{dgammaC}
\end{equation}
$B_2^\prime$ is given in Eq.~(7) of Ref.~\cite{dos}. The infrared convergent $I_{C0F}$, explicitly given in Eq.~(37) of
Ref.~\cite{cinco}, corresponds to the infrared divergent $I_{C0}$ of the TBR. $C_{AF}^{(s)}$ can be arranged as the sum
\begin{equation}
C_{AF}^{(s)}=C_{IF}+C_{IIF}+C_{IIIF},
\end{equation}
where
\begin{equation}
C_{IF}=\sum_{i=1}^{8}Q_{i+6}\Lambda_{iF},\qquad C_{IIF}=\sum_{i=6}^{15}Q_i\Lambda_{(i+3)F}, \qquad
C_{IIIF}=\sum_{i=16}^{25}Q_i\Lambda_{(i+3)F}.
\end{equation}
which are the counterparts of Eqs.~(21)-(23) of Ref.~\cite{dos}.

The $Q_i$ $(i=6,\ldots,25)$ are quadratic functions of the form factors and are common to both regions. Their explicit expressions are
found in Appendix A of Ref.~\cite{dos}. It should be clear that for $Q_6$ and $Q_7$ we use $\tilde{Q}_6$ and $\tilde{Q}_7$ of this
Appendix, since the contributions of orders $(q/M_1)^2$ and higher have been subtracted.\footnote{In Ref.~\cite{dos}
$\tilde{Q}_9$ actually corresponds to $Q_9$ due to the order of approximation implemented.}

The analytical form of $C_{AF}^{(s)}$ is obtained by performing explicitly the triple integrals over the real photon variables,
contained in the $\Lambda_{iF}$ functions ($i=1,\ldots,28$). These integrals result in a set of analytical functions $\theta_{jF}$
($j=0,\ldots,22$). The connection between the $\Lambda_{iF}$ and the $\theta_{jF}$ functions requires several algebraic steps, which
are exhibited in terms of intermediate functions $X_{iF}$, $Y_{iF}$, $Z_{iF}$, $N_{iF}$, $I_{F}$, $\eta_{0F}$, $\gamma_{0F}$,
$\chi_{ijF}$, and $\zeta_{ijF}$.

Once the final results for the FBR are available it is only necessary to give explicitly the $\theta_{jF}$ functions to present it. The
$\Lambda_{iF}$ and the intermediate functions\footnote{In Ref.~\cite{dos} we have detected a misprint in the expression
of the intermediate function $\mathcal{J}$, defined at the end of Appendix D of this reference. In the term proportional to
$\theta_{12}$, the denominator of the first summand should read $\beta^2$ rather than $\beta$. This misprint, however,
does not affect any of the results or conclusions.} can be obtained from the corresponding $\Lambda_i$ and $X_i$, $Y_i$, $Z_i$, $N_i$, $I$,
$\eta_0$, $\gamma_0$, $\chi_{ij}$, and $\zeta_{ij}$, respectively, of the TBR by making some simple changes in the latter. This is
possible because of the parallelism mentioned above. These changes are: (1) a subindex $F$ is attached to all the corresponding
functions of the TBR, (2) the terms proportional to the factor $(1-y_0)$ are replaced by zero, and (3) otherwise the factor $y_0$ is
kept as such. There are three exceptions to rule (3). In $I$, $\chi_{11}$, and $\eta_0$, $y_0$ appears by itself only once and there it
must be replaced by $y_0=1$ to produce $I_{F}$, $\chi_{11F}$, and $\eta_{0F}$ respectively. The $\Lambda_i$, $X_i$, $Y_i$, $Z_i$,
$N_i$, $I$, $\eta_0$, and $\gamma_0$ are found in Appendix B of Ref.~\cite{dos}. It should be stressed that
$\Lambda_{2F}=0$ as is $\Lambda_2=0$, and that is why it does not appear in this Appendix. The $\chi_{ij}$, and $\zeta_{ij}$ are
found in Sec.~IV of Ref.~\cite{seis}, but $\zeta_{10}$, $\chi_{10}$, and $\chi_{20}$ were not directly given in terms of $\theta_i$
there and the above rules cannot be easily applied to them. It is better to give the explicit connection now, namely,
\[
\zeta_{10F} = p_2l \left[ \theta_{13F}-\frac{l}{p_2}\theta_{10F}-\frac{E_\nu^0}{p_2} \theta_{5F}\right],
\]
\[
\chi_{10F} = l\left[ 2\eta_{0F}-E_\nu^0 \theta_{4F} - l\theta_{5F} \right],
\]
\[
\chi_{20F} = 2l^2\left[ p_2y_0 \theta_{4F} + E_\nu^0 \theta_{5F} + 2l\theta_{10F}-p_2\theta_{13F}-\frac12 \theta_{14F}
- \frac{E_\nu^0}{l}\eta_{0F}\right].
\]
The function $\zeta_{31}$ of Ref.~\cite{dos} has also a misprint in the term proportional to $\theta_{12}$. In order to
avoid further confussions, we provide here the right expression for $\zeta_{31F}$, which reads
\begin{eqnarray}
\zeta_{31F} & = & p_2ly_0 \left[ 2(3E^2-l^2) \theta_{3F}-6E^2(\theta_{4F}+\beta \theta_{5F}) + \theta_{9F} \right]
- 30lE^2p_2\theta_{13F} - 30l^2Ep_2 \theta_{19F} \nonumber \\
&  & \mbox{} -\frac{6l^3}{\beta^4} \left[ 5(l + \beta E_\nu^0) +
3\beta^2(p_2y_0-l) \right] (\theta_{3F}-\theta_{4F}-\beta
\theta_{5F}) - 18l^2EE_\nu^0(\theta_{4F}-\theta_{3F}) \nonumber \\
&  & \mbox{} + 6p_2l^3y_0 \theta_{3F} + 30lE^2 (l + \beta E_\nu^0)
\theta_{10F} + 30El^3 \theta_{20F} - \frac12 \theta_{22F}
\nonumber \\
&  & \mbox{} -6p_2\left[E^2l(\beta^2-5) - \frac{2lp_2^2+2\beta
p_2ly_0(E+E_\nu^0)}{b^+b^-}\right] \theta_{12F}, \nonumber
\end{eqnarray}
so that the correct $\zeta_{31}$ of the TBR can be obtained by dropping the subindex $F$ from the $\theta$'s in the above
equation. We need to point out that neither results nor conclusions are affected by that misprint.

Let us illustrate the above procedure with an example. $\Lambda_{15F}$ is obtained from
\begin{eqnarray}
\Lambda_{15} & = & \frac{\beta(E+E_\nu^0)}{4M_1} \left[ p_2^2\theta_7 + 2p_2^2 E(\theta_4-\theta_3)+2\zeta_{21}
- \frac{2}{l}X_3\right] + \frac{X_{4}}{2M_1} \nonumber \\
&  & \mbox{} -\frac{l^2p_2}{4M_1} (y_0^2-1) + \frac{p_2}{4M_1E}\left[ 4(ly_0+p_2)X_2 + p_2\chi_{21}\right], \nonumber
\end{eqnarray}
by first putting $y_0=1$ into the factor $(1-y_0)$ only and keeping $y_0$ as such elsewhere, and second by attaching a subindex $F$ to
the intermediate functions. The result is
\begin{eqnarray}
\Lambda_{15F} & = & \frac{\beta(E+E_\nu^0)}{4M_1} \left[ p_2^2\theta_{7F} + 2p_2^2E(\theta_{4F}-\theta_{3F}) + 2\zeta_{21F}
- \frac{2}{l}X_{3F}\right] + \frac{X_{4F}}{2M_1} \nonumber \\
&  & \mbox{} + \frac{p_2}{4M_1E} \left[ 4(ly_0+p_2) X_{2F} + p_2\chi_{21F}\right]. \nonumber
\end{eqnarray}
With these changes all we now need is the list of the $\theta_{jF}$ functions. Their explicit forms for $j=1,\ldots,16$ were already
calculated in Ref.~\cite{cinco} and they are found in Appendix B of this reference. $\theta_{0F}$ is given in Eq.~(38) of this
reference, $\theta_{1F}=\theta_{17F}=0$, and $\theta_{18F}=1$.

The functions $\theta_{jF}$ with $j=19,\ldots,22$ are new. They are defined as $\theta_{19F}=\int_{-1}^1 x \xi_{4}^{T}(x) dx$,
$\theta_{20F}=\int_{-1}^{1}x^{3}\xi_1^{T}(x) dx$, $\theta_{21F}=\int_{-1}^{1}x^2\xi_2^{T}(x) dx$, and $\theta_{22F}=\int_{-1}^{1}
\xi_{6}^{T}(x)/(1-\beta x)dx$. The functions $\xi_1^{T}(x)$, $\xi_2^{T}(x)$, and $\xi_{4}^{T}(x)$ are found in the Appendix of
Ref.~\cite{tres}. Here we only need the explicit form of $\xi_{6}^{T}(x)$, which reads
\begin{equation}
\xi_{6}^{T}(x) = h_1(x)(1-\beta x) + \frac{h_2(x)}{(x+a^{+})(x+a^{-})} + \frac{h_3^{+}(x)}{(x+a^{+})^2} + \frac{h_3^{-}(x)}{(x+a^{-})^2}
+ \frac{h_{4}^{+}(x)}{(x+a^{+})} + \frac{h_{4}^{-}(x)}{(x+a^{-})},
\end{equation}
where

\begin{subequations}
\label{eq:hi}
\begin{eqnarray}
h_1(x) & = & 24p_2l^2\left[ \frac{a^{+}y_0^{+}}{b^{+}(x+a^{+})} + \frac{a^{-}y_0^{-}}{b^{-}(x+a^{-})}\right] x, \\
h_2(x) &= & 8l\left\{ 6x\left[ 2a_1^2+p_2^2-y_0p_2xa_1\right] - p_2^2x(3+x^2) - 3y_0p_2a_1(1+x^2) + 6xa_1^2\right. \nonumber \\
&  & \left. + 6p_2\left[ 2y_0a_1- (y_0^2+1) p_2x \right]\right\},\\
h_3^\pm(x) & = & 2l \left\{ 2p_2^2x(3+x^2) + 6y_0p_2a_1(1+x^2) - 12xa_1^2\mp 2p_2^2y_0(y_0^2+3) \right. \nonumber \\
&  & \left. \pm 6a_1\left[ 2y_0a_1- (y_0^2+1) p_2x\right] \right\}, \\
h_{4}^\pm(x) & = & \mp \left[\frac{4l^2}{p_2}\right] \left[ 3p_2(p_2x+y_0a_1) - a^\pm(p_2^2x^2+3y_0p_2a_1x-6a_1^2) \right].
\end{eqnarray}
\end{subequations}
We have used the definitions
\begin{equation*}
a^\pm = \frac{E_\nu^0 \pm p_2}{l}, \qquad b^\pm = 1+\beta a^\pm, \qquad y_0^\pm = y_0 \pm a^\pm, \qquad a_1=E_\nu^0+lx.
\end{equation*}

The analytical results of $\theta_{19F}$, $\theta_{20F}$, $\theta_{21F}$ and $\theta_{22F}$ are
\begin{equation*}
\theta_{19F}= \frac{4}{3p_2}, \qquad \theta_{20F}=\frac{1}{p_2} (T_{20F}^{+}+T_{20F}^{-}), \qquad \theta_{21F} =T_{21F}^{+}+T_{21F}^{-},
\qquad \theta_{22F} = T_{22F}^{+}+T_{22F}^{-},
\end{equation*}
where
\begin{eqnarray}
T_{20F}^\pm & = & \pm \frac{1}{4}\left[ 1-(a^\pm)^{4}\right] I_2^\pm\pm \frac{a^\pm}{2}\left[ (a^\pm)^2+\frac{1}{3}\right], \nonumber \\
T_{21F}^\pm & = & \frac43 \mp 2a^\pm(y_0 \pm a^\pm) \left[a^\pm I_2^\pm-2\right], \nonumber \\
T_{22F}^\pm & = & \frac12 L_0 + \frac12 L_1I_1 + L_2^\pm I_2^\pm + L_3^\pm I_3^\pm. \nonumber
\end{eqnarray}
The $I_1$, $I_2^\pm$, and $I_3^\pm$ functions are found at the end of Appendix C of Ref.~\cite{seis} and the $L_i$ are defined as
\begin{eqnarray}
L_0 & = & 48l^2p_2(\frac{a^{+}y_0^{+}}{b^{+}}+\frac{a^{-}y_0^{-}}{b^{-}}) -16E\left[p_2^2 +3l(2l-p_2y_0) \right], \nonumber \\
L_1 & = & \frac{\beta^2h_2(1/\beta)}{b^{+}b^{-}} + \frac{\beta \left[ \beta h_3^{+}(1/\beta) + b^{+}h_{4}^{+}
(1/\beta) \right]}{(b^{+})^2} + \frac{\beta \left[ \beta h_3^{-}(1/\beta) + b^{-}h_{4}^{-}(1/\beta) \right]}
{( b^{-})^2}, \nonumber \\
L_2^\pm & = & \frac{-24p_2l^2(a^\pm)^2y_0^\pm + h_4^\pm(-a^\pm)}{b^\pm} + \frac{b^\pm h_3^\pm(0) - h_3^\pm(-a^\pm)}{b^\pm a^\pm}
- 4Ea^\pm\left[ 3l(2l-y_0p_2) - p_2^2\right] \nonumber \\
&  & \mbox{} - \frac{\beta^2a^\pm h_3^\pm(1/\beta)}{(b^\pm)^2}\mp \frac{lh_2(-a^\pm)}{2p_2b^\pm}, \nonumber \\
L_3^\pm & = & \frac{h_3^\pm(-a^\pm)}{b^\pm}, \nonumber
\end{eqnarray}
where the functions $h_i(x) $ are the same of Eq.~(\ref{eq:hi}), but evaluated at the indicated arguments.

For a neutral process we proceed again in parallel to the calculation of $d\Gamma_{\textrm{N}B}^{(s)}$ given in Sec.~IIIB of
Ref.~\cite{dos}. Thus, we can express $d\Gamma_{\textrm{N}B}^{(s)\, \textrm{FBR}}$ as in Eq.~(31) of this reference as
\begin{equation}
d\Gamma_{\textrm{N}B}^{(s)\, \textrm{FBR}} = \frac{\alpha}{\pi} d\Omega {\hat {\mathbf s}_1} \cdot {\hat {\mathbf p}_2} \left[
B_2^\prime I_{N0F} + C_{AF}^{(s)} + C_{NAF}^{(s)}\right],
\end{equation}
where $B_2^\prime$ and $C_{AF}^{(s)}$ are those of Eq.~(\ref{dgammaC}). The function $I_{N0F}$ corresponds to the infrared divergent
$I_{N0}$ of Eq.~(35) of Ref.~\cite{siete}. As in the charged case, $I_{N0F}$ is no longer infrared divergent. It can be calculated up
to order $(q/M_1)$ by using the approximation $1/p_2\cdot k\simeq 1/p_1\cdot k+q\cdot k/(p_1\cdot k)^2$. The result is
\[
I_{N0F} = I_{C0F} + \delta I_0,
\]
where $I_{C0F}$ is the same of Eq.~(\ref{dgammaC}) and
\[
\delta I_0 = \frac{4p_2}{M_1\beta}\left[ \frac{y_0}{2}\ln\left(\frac{y_0+1}{y_0-1}\right) - 1\right] \left[ \left(
\frac{1-\beta^2}{\beta}\right) \textrm{arctanh}\, \beta - 1\right].
\]
It should be remarked that this expression is equal to zero when $\beta
\rightarrow 0$ and $y_0\rightarrow \infty .$

The other new integrals appear only in $C_{NAF}^{(s)}$. They are
\[
C_{NAF}^{(s)}= D_3\rho_{N3F}+D_{4}\rho_{N4F}.
\]
Here $D_3=2(f_1g_1-g_1^2)$ and $D_4=2(f_1g_1+g_1^2)$, $\rho_{N3F}$ and $\rho_{N4F}$ are
\begin{eqnarray}
\rho_{N3F} & = & \rho_{IF} + \rho_{IIF} + \rho_{IIIF}, \nonumber \\
\rho_{N4F} & = & \rho_{IF}^\prime + \rho_{IIF}^\prime + \rho_{IIIF}^\prime. \nonumber
\end{eqnarray}
The analytical forms of the integrals $\rho_{iF}$ and $\rho_{iF}^\prime$ ($i=I,II,III$) can be obtained from Eqs.~(34)-(39) of
Ref.~\cite{dos} again by changing $y_0$ to one as the upper limit in the $y$ integrals. After the integration one can see that the
analytical form of the $\rho_{iF}$ and $\rho_{iF}^\prime$ can be obtained from the analytical expressions of $\rho_i$ and
$\rho_i^\prime$ for TBR, which are found in Appendix C of Ref.~\cite{dos}, using the same rules (1), (2), and (3) applied to
$\Lambda_i$ to obtain the $\Lambda_{iF}$ of the FBR for charged decaying baryons.

We have made crosschecks between numerical integrals and analytical results of the $\Lambda_{iF}$, $\rho_{iF}$, and $\rho_{iF}^\prime$,
and they were satisfactory. Our results are model-independent and include terms up to order $(\alpha/\pi)(q/M_1)$, this is as far one
can improve the precision of RC before introducing model-dependence into them. They are useful when $l=e^\pm,\mu^\pm$, and $\tau^\pm$.
The two cases discussed here, charged and neutral decays, allow one to cover the other four charge assignments predicted by heavy
quarks in baryons \cite{ocho}. When such quarks are involved, the results of Ref.~\cite{uno} are useful in low statistics experiments
(several hundreds of events) and the improved results presented here are useful in medium statistics experiments (several thousands of
events).

The authors acknowledge financial support from CONACYT, COFAA-IPN, and FAI-UASLP (Mexico).

\end{document}